\documentclass[epj]{svjour}
\usepackage{amssymb}
\usepackage{graphics}
\begin{document}
\title{Melting of a Wigner Crystal in an Ionic Dielectric}
\author{S. Fratini \and P. Qu\'{e}merais 
}                     
\offprints{quemerai@lepes.polycnrs-gre.fr}          
%
\institute{Laboratoire d'Etudes des Propri\'{e}t\'{e}s Electroniques des
Solides\\
CNRS, BP 166, 38042 Grenoble Cedex 9, France}
\date{Received: date / Revised version: date}
%
%

\abstract{
The melting of a Wigner Crystal of electrons placed into a host polar
material is examined as a function of the density and the temperature.
When the coupling to the longitudinal optical modes of the host medium is
turned on, the WC is progressively transformed into a polaronic Wigner
Crystal.  We estimate the critical density for crystal melting at zero
temperature using the Lindeman criterion.  We show that above a certain
critical value of the Fr\"ohlich electron-phonon coupling, the melting
towards a quantum liquid of polarons is not possible, and the
insulator-to-metal transition is driven by the ionization of the polarons
(polaron dissociation).  The phase diagram at finite temperature is
obtained by making use of the same Lindeman criterion.  Results are also
provided in the case of an anisotropic electron band mass, showing that the
scenario of polaron dissociation can be relevant in anisotropic compounds
such as the superconducting cuprates at rather moderate e-ph couplings.
\PACS{
      {PACS-key}{describing text of that key}   \and
      {PACS-key}{describing text of that key}
     } 
} 
\maketitle

\section{Introduction}

\label{sec:Introduction}

As it is well-known,  the formation of a
single polaron is already a many-body
problem, since it deals with the interaction of one electron with an
infinite number of phonons. It is thus clear that treating a system of
polarons at finite density is a very complicated task. Some
theories have been proposed so far which neglect
the Coulomb interactions, or treat them in  the random phase
approximation. The very existence of polaronic bands \cite{alexandrov}
lies on
this theoretical basis, although, to our knowledge, no rigorous
justification of this approach has ever been given
in the literature. Old but reliable results were obtained by
Mahan for low electron-phonon (e-ph) coupling and high electron density
\cite{mahan}, and extended
later to intermediate e-ph coupling by Lemmens et al. \cite{lemmens}. Recently,
De Filippis et al. \cite{iadonisi} obtained interesting results
valid at lower densities and higher
e-ph coupling, giving a more complete physical insight to the problem in
the metallic phase.

A completely different approach of the
same problem can be given starting from the
low density regime, where the particles localize due to the long-range Coulomb
repulsion \cite{wigner,quem1,remova}. In this limit, the treatment is
somehow simplified because the
statistical effects between the electrons can be neglected.
The simplest way to
introduce the problem we are dealing with is to consider a
Wigner Crystal (WC) of electrons, which is the ground-state of the
electron gas at low density, and dip it into a polar (ionic) host
material. What are the effects of the polarization on the
stability of the WC?

A polar dielectric is a compound which is characterized by
two main sources of polarization: the electronic
polarization, corresponding
to the vibrations of the core electrons at optical frequencies
($10^{14-15} s^{-1}$),  and the ionic
polarization, which is carried by the
longitudinal optical modes in the infra-red region ($\omega_{LO} \approx
10^{13} s^{-1}$), corresponding to the relative motion of the positive
and negative ions.
An external charge moving in such material \cite{remark-external} interacts
with the
dipoles of the dielectric, and produces a polarization cloud which
partially screens its Coulomb field. The composite state  of an electron plus
its induced polarization is called a \textit{polaron} \cite{landau}.
Such a quasi-particle has a finite
extension (the polaron radius), a finite self-energy (the energy required
to form the bound state), and a renormalized mass $M_P$ which is generally
greater than the electron band mass.
As customary and following F\"rohlich \cite{frohlich},
we can define a dimensionless e-ph coupling as $\alpha =\left( {{{m^{*}%
}/{2\hbar ^{3}\omega _{LO}}}}\right) ^{1/2}{{e^{2}}/{\tilde{\varepsilon}}}$,
which is the only relevant parameter for a single isolated polaron. Here $m^*$
is the electron band mass
and the effective dielectric constant $\tilde{\varepsilon}$ is
 defined as $1/\tilde{\varepsilon}=1/\varepsilon _{\infty
}-1/\varepsilon _{s}$, where $\varepsilon_s$ and $\varepsilon_\infty$
are respectively the static and
high frequency dielectric constants of the medium.

If we now consider two interacting polarons, two different situations can
occur.  Under some very specific conditions, the two particles can attract
and form a bound state, sharing the same polarization cloud.  The
properties of such composite bosons --- the \textit{bipolarons} --- have
been extensively studied, especially in connection to their Bose
condensation at low enough temperatures.
However, bipolarons can only exist if the ratio
$\eta=\varepsilon _\infty/\varepsilon _s$ is lower than a critical value
$\eta_c \approx 0.1$, and for rather
large e-ph couplings ($\alpha> 6-7$)  \cite{bipolarons}. On the other hand, for
$\eta>\eta_c$, which is quite common in polar materials, two polarons
repel at large distances as $1/\varepsilon_s d$. Throughout the present
paper, we
will be dealing with compounds where the net interaction between polarons
is repulsive.

As was stated previously, a reliable treatment of the many-polaron
problem can be given starting at low density, if we assume that the
particles localize due to the long-range Coulomb interactions, forming a
Polaronic Wig\-ner Crystal (PWC).
In that case, the polarization of the host material will react to screen the
external charges, and this will change both their interaction and kinetic
energy, as compared to the ordinary WC of electrons.  On one hand,
the total Coulomb interaction energy is
shifted from $\sim e^2 / R_s$ to $\sim e^2 / \varepsilon_s R_s$,
$R_s$ being the mean distance between two electrons at a
given density.
On the other hand, the
kinetic energy is also reduced owing to the polaron mass, and behaves as
$\sim \hbar^2 / M_p R_s^2$. From this simple argument,
we see that the two effects above compete regarding the WC stability. While the
decrease of the interaction energy favours delocalization,
 the decrease of the kinetic energy tends to stabilize the
crystallized state.

The correct evaluation of the balance between these two effects
is thus essential in order to understand the melting mechanism occurring
for increasing density.
It is particularly instructive to consider the adiabatic limit obtained
for vanishing phonon frequency $\omega_{LO}$
($\alpha\rightarrow \infty$). In that case, as given by standard polaron
theory \cite{allcock}, while the polaron radius and energy remain finite,
the polaron mass becomes infinite, and there is no way to get mobile
polarons. Therefore, at
zero temperature, the PWC can only melt  through ionization of
the electrons from their
polarization potential-wells. The latter must be  screened out by the
liberated carriers in the same fashion
as for the usual Mott transition in non-polar semiconductors, and the
system will eventually become metallic \cite{mott}. We have already
published a phenomenological Mott-like criterion which
describes the insulator-to-metal transition (IMT) in this situation
\cite{quem2}:
\begin{equation}
n_{c}^{1/3}\left( {\tilde{\varepsilon}}/\varepsilon _{\infty }\right)
R_{P}\approx 0.25  \label{Mott-gen}
\end{equation}
where $R_{P}$ is the polaron radius (the bound state radius), which is the
relevant localization length in our problem.

If one restores a finite phonon frequency ($\omega_{LO} \neq 0$) as is
the case in real materials, the problem becomes more
intricate, since the polarons are \textit{a priori}
mobile (they have a finite mass). Upon increasing the density, the
polarons themselves could be delocalized without being ionized, possibly
leading to a polaron liquid beyond the transition.
The aim of the present work  is to
examine carefully the quantum melting of a PWC in all the e-ph coupling
regimes, as was previously done for the more simple case of the electron
crystal
(which is recovered for $\alpha=0$).
Some important results have already been published
in several letters \cite{quem1,quem2,frat1}. We provide here  the
details concerning the model, approximations and calculations, and give a
more complete description including
 results at finite temperatures and for anisotropic compounds \cite{FratPhD}.

The paper is organized as follows.  In section 2, we derive a model for the
insulating state which includes both electron-phonon and electron-electron
interactions, and point out the different approximations involved.
Starting from the Fr\"ohlich model, we assume a crystallized state for the
polarons, and expand the long-range Coulomb interactions between different
particles up to second order in the small displacements relative to their
equilibrium positions, neglecting anharmonicity and exchange effects.
It can be demonstrated \cite{frat1,FratPhD} that the resulting
dipole-dipole interactions between different electrons  are responsible
for the dispersion of the collective modes of the
 PWC. Furthermore,  above a certain critical density,
the dipolar terms can lead to a phonon instability of the polaron
lattice, corresponding to the spontaneous excitation of long-wavelength
transverse modes. This phenomenon, as well as its consequences on the
properties
of the dielectric constant of the system, will be extensively analysed
in a forthcoming paper \cite{frat2}.

In the present work, we will neglect such non-local interaction terms,
restricting ourselves to the mean-field Wigner approximation.  This
corresponds to including the many-body effects through an effective local
potential acting on each particle. Hence, the many-polaron problem is
reduced to the problem of a single polaron in an external potential.
Taking advantage of this approximation, in section 3 we study the
  PWC in the framework of Feynman's path-integral method
\cite{feynman}, which is known to give reliable analytical results at any
values of the e-ph coupling.

In section 4,
we examine the insulator-to-metal transition occurring for increasing
density (at zero temperature) in terms of the Lindemann criterion, as was done
by Nozi\`{e}res and Pines
 for the electron crystal \cite{nozieres}. Since the polaron is a
composite particle, we argue that there are two different Lindemann
criteria to be
considered: one for the center-of-mass motion, which describes the
melting as in the ordinary WC, and one for the relative
displacement between the electron and its polarization cloud, which describes
the ionization of the polarons. This allows us to
distinguish between two differernt melting mechanisms, depending on the e-ph
coupling strength.
At weak or moderate couplings ($\alpha<\alpha^*$), the situation
is very much the same as in the bare electron crystal, i.e. the
melting is driven by the increasing fluctuations of the localized
particles.
On the other hand, at
strong e-ph coupling ($\alpha>\alpha^*$), i.e. at low
but finite $\omega _{LO}$, the crystal melting is due to the
dissociation of the polarons. In that case, the electrons cannot carry
with them their phonon clouds in the metallic phase, and the melting
towards a degenerate polaron liquid can be excluded.
We next extend the results to finite temperatures, and derive an
approximate $T\; vs. \; n$ phase diagram for the many-polaron problem. For
an application to the
superconducting cuprates,
we finally generalize our treatment to
systems where the electron band mass is an anisotropic quantity. The
results presented in Table 1 suggest that polaron
dissociation could be a real physical possibility in such compounds.

\section{Model and approximations}

\label{sec:modele}

In this  section, we derive our basic model for the polaron crystal.
Let us consider a polar material,
characterized by the three parameters aforementioned: $\omega _{LO}$,
$\varepsilon _{s}$ and $\varepsilon_{\infty }$. We next consider an
ideal doping procedure, which consists in introducing into this "host"
material a finite density
of electrons, which we express as $n=(4\pi R_s^3/3)^{-1}$, plus a {\it
rigid} jellium which
exactly compensates the negative charges.
Owing to its polarizability, the
host crystal responds to these \textit{excess} charges. If we neglect
magnetic effects,  the total interaction energy
is given by standard laws of electrostatics as \cite{allcock}:
\begin{equation}
{\cal E}_{I}=\frac{1}{4\pi }\int \int {\bf {\ E}}\cdot \dot{{\bf {\ D}}}%
\,d^{3}r\,dt  \label{E_em}
\end{equation}
where time integrations  must be carried out along a path
satisfying the equations of motion.
If we suppose that the localization length scales of our problem are
larger than the anion-cation distance in the host material,
the latter can be treated as a continuous  medium,
and we can  consider the total electric field
${\bf E}({\bf r})$ and the electric
displacement ${\bf D}({\bf r})$ due to the  doping charges
(electrons plus jellium)
as  macroscopic fields. Such a treatment is standard in
polaron theory, and leads to the well-known Landau-Fr\"ohlich model \cite
{landau,frohlich}.
The two quantities  above defined are related to
the polarization field
${\bf P}$  through the relation
${\bf D}={\bf E}+4\pi {\bf P}$, which  can be used to eliminate
 ${\bf E}$, and derive a Lagrangian which  only depends on the
excess charges. Following Fr\"ohlich, we separate two sources
of polarization: ${\bf P}={\bf P}_{0}+{\bf P}_{ir}$. ${\bf P}_{0}$
corresponds to
high frequency
oscillating charges
and can be
included through the optical dielectric constant $\varepsilon_\infty$.
${\bf P}_{ir}$ is due to the ionic distortion, and vibrates
at a frequency $\omega _{LO}$  in the infra-red range.
One obtains the following interaction
 Lagrangian  (see ref. \cite{allcock} for more details about this
derivation):
\begin{equation}
L_{I}=-\frac{1}{8\pi \varepsilon _{\infty }}\int {\bf {\ D}}^{2}\,+\int {\bf
{\ P}}_{ir}\cdot {\bf {\ D}}\,+\frac{2\pi \tilde{\varepsilon}}{\omega
_{LO}^{2}}\int \left( \dot{P_{ir}}^{2}-\omega _{LO}^{2}P_{ir}^{2}\right)
\label{L-int}
\end{equation}
where ${\bf {\ P}}_{ir}$ and $(4\pi \tilde{\varepsilon}/\omega
_{LO}^{2})\dot{{\bf {\ P}}}_{ir}$ are conjugate moments, and integrations
run over the entire space ($d^{3}{\bf r}$ in the integrands
has been omitted). The different terms in eq. (\ref{L-int})
represent respectively the interactions between the excess charges,
between the charges and the longitudinal optical phonons, and the
free-phonon Lagrangian. We can
now specify the nature of the doping charges --- say electrons
and positive jellium, but the same applies to holes and negative jellium ---
by writing ${\bf {\ D}}={\bf {\ D}}^{-}+{\bf {\ D}}^{+}$, where
\begin{equation}
{\bf {\ D}}^{-}=\sum_{i}{\bf {\ D}}_{i}^{-}=-e\sum_{i}{\frac{{\bf r}-{\bf r}%
_{i}}{|{\bf r}-{\bf r}_{i}|^{3}},}
\end{equation}
${\bf r}_{i}$ being the electron coordinate. We will come back later on the
precise formulation of the term ${\bf D}^{+}$ due to the rigid jellium.

Separating  explicitely the two parts of the electric displacement, and
adding the kinetic energy of the free electrons, one gets the
following many-body Lagrangian:
\begin{eqnarray}
\lefteqn{L=\sum_{i}\frac{m^{*}}{2}\dot{r}_{i}^{2}\,
+\frac{2\pi \tilde{\varepsilon}}{\omega _{LO}^{2}}\int
\left( \dot{P}_{ir}^{2}-\omega _{LO}^{2}{P}_{ir}^{2}\right) } \label{L_tot} \\
&&-\frac{1}{8\pi \varepsilon _{
\infty }}\int \mathbf{D}^{+}\cdot \mathbf{D}^{+}
-\frac{1}{8\pi \varepsilon _{\infty }}
\int \sum_{i\neq j}\mathbf{D}_{i}^{-}\cdot \mathbf{D}_{j}^{-}  \nonumber \\
&&-\frac{1}{4\pi \varepsilon _{\infty }}\int {\mathbf {D}}^{+}\cdot (\sum_{i}
{\bf {D}}_{i}^{-})\,+\int \mathbf{P}_{ir}\cdot (\sum_{i}{\bf {D}}_{i}^{-}+
{\bf {D}}^{+})\, \nonumber
\end{eqnarray}
The interacting terms in (\ref{L_tot}) now correspond  to the self-energy of
the jellium, the Coulomb interactions among different electrons, the
electron-jellium,  electron-phonon and jellium-phonon
interactions.  Since the jellium is assumed to be rigid,
it is convenient to
introduce the polarization field which only responds to the electron motion by
making use of the following transformation:
\begin{equation}
{\bf \tilde{P}}_{ir}={\bf P}_{ir}-\frac{1}{4\pi {\tilde{\varepsilon}}}{\bf D}%
^{+}
\end{equation}
so that (\ref{L_tot}) becomes:
\begin{eqnarray}
\lefteqn{L=\sum_{i}\frac{m^{*}}{2}\dot{r}_{i}^{2}\,
+\frac{2\pi \tilde{\varepsilon}}{\omega _{LO}^{2}}\int \left(
\tilde{P}_{ir}^{2}-\omega _{LO}^{2}{\tilde{P}}_{ir}^{2}\right) }\label{Lfinal}\\
&&-\frac{1}{8\pi \varepsilon _{s}}%
\int \mathbf{D}^{+}\cdot \mathbf{D}^{+}-\frac{1}{8\pi \varepsilon _{\infty
}}\int
\sum_{i\neq j}\mathbf{D}_{i}^{-}\cdot \mathbf{D}_{j}^{-}  \nonumber \\
&&-\frac{1}{4\pi \varepsilon _{s}}\int {\mathbf{D}}^{+}\cdot
(\sum_{i}\,{\mathbf {D}}_{i}^{-})\,+\int {\tilde{{\bf {P}}}}_{ir}
\cdot (\sum_{i}{\bf {D}}%
_{i}^{-})\, \nonumber
\end{eqnarray}
(the phonon-jellium interaction has disappeared, and the remaining terms
involving
the jellium are now divided by $\varepsilon_s$).
The many-body Lagrangian in its final form (\ref{Lfinal})
depends on the following parameters: (i) the electron density $n$; (ii)
the  parameters of the host
material $m^{*}, \varepsilon _{s},\varepsilon _{\infty },\omega _{LO}$, which
determine the e-ph coupling $\alpha$.

\subsection{Phonon integration}
\label{sec:phint}

The exact partition function of the system can be
expressed in the
path integral formulation as:
\begin{equation}
Z=Tr(e^{-\beta H})=\int e^{S} \; {\cal D}(path)  \label{Z}
\end{equation}
where  path integrals run over all the electron coordinates and on the
polarization field. $S$ is defined as usual as the  integral of $L$ in
imaginary time ($t=i\tau $, $\beta ^{-1}$ being the
temperature, and taking $\hbar=k_B=1$). The trace operation
corresponds to integrating over all possible closed trajectories.
Since the terms depending on the polarization are quadratic,
the field $\tilde{\mathbf{P}}_{ir}$ can be exactly
integrated out \cite{feynman}. As a result, the
total effective action for the electrons is the sum of three terms:
\begin{equation}
S(\{{\bf r}_{i}\})=S_{jellium}+S_{e}(\{{\bf r}_{i}\})+S_{e-ph}(\{{\bf r}%
_{i}\})  \label{S total}
\end{equation}
The first term is constant at fixed density and does not depend on the state
of the system:
\begin{equation}
S_{jellium}=-\frac{\beta }{8\pi \varepsilon _{s}}\int{{\mathbf {D}}^{+}%
\cdot{\mathbf {D}}^{+}}  \label{Sjellium}
\end{equation}
The second term in (\ref{S total}) is the electronic part:
\begin{eqnarray}
S_{e}(\{{\bf r}_{i}\})&=&\sum_{i}\frac{m^*}{2} \int_0^\beta
\dot{r}_{i}^{2}d\tau-\frac{%
1}{8\pi \varepsilon _{\infty }} \int_0^\beta \int\sum_{i\neq j}{\bf
{D}}_{i}^{-}\cdot
{\bf {D}}_{j}^{-} d\tau
 \nonumber\\
&-& \frac{1}{4\pi %
\varepsilon _{s}} \int_0^\beta \int
{\bf {D}}^{+}\cdot (\sum_{i}\,{\bf {D}}_{i}^{-}) d\tau \label{Se}
\end{eqnarray}
It contains the electron kinetic energy, the instantaneous Coulomb repulsion
between the electrons, and the interaction between the
electrons and the jellium.

The electron-phonon coupling effects are included in $S_{e-ph}$, which is
given by:
\begin{equation}
S_{e-ph}(\{\mathbf{r}_{i}\})=\sum_{i,j}{\frac{\omega _{LO}e^{2}}
{4{\tilde{\varepsilon}%
}}\int_{0}^{\beta } \! {\int_{0}^{\beta }{\frac{G_{\omega _{LO}}(\beta
,\tau -\sigma )}{|{\bf r}_{i}(\tau )-{\bf r}_{j}(\sigma )|}}d\tau d\sigma }%
}  \label{Se-ph}
\end{equation}
We have introduced the phonon propagator
$\mbox{$G_{\omega }(\beta ,\tau -\sigma )$}=(\bar{n}%
+1)e^{-\omega |\tau -\sigma |}+\bar{n}e^{\omega |\tau -\sigma |}$,  together
with $\bar{n}=(e^{\beta \omega }-1)^{-1}$.
Formula (\ref{Se-ph}) represents the {\it retarded} interactions between
electrons,
mediated by the lattice polarization: the diagonal terms ($i=j$) correspond
to the
interaction of each
electron with itself, i.e. the polaron effect, while the off-diagonal terms (%
$i\ne j$) give a retarded attraction between electrons $i$ and
$j$.

The above expressions (\ref{S total})--(\ref{Se-ph}) are valid at any
temperature.
We will now apply our hypothesis of crystallization at low densities, and
make use of several approximations.

\subsection{Model for the Crystallized State}

Let us first consider the jellium. We will assume from now on that it is
constitued of spheres of radius $R_s$ with a uniform positive charge density
$\rho^+=e/\left( 4\pi R_{s}^{3}/3\right) $.  Each sphere, which carries a total
 charge $+e$, is supposed to be centered on the sites $%
\left\{ {\bf R}_{i}\right\} $ of a
Bravais lattice. This approximation is intermediate between
considering actual doping ion potentials, and a jellium uniformly spread
out in the system.
There is a small overlap between adjacent spheres, that we will neglect
in the evaluation of the electrostatic energy. This is
convenient to evaluate the jellium-jellium and the electron-jellium
interaction terms.
In particular, one can write:
\begin{equation}
{\bf D}^{+}\left( {\bf r}\right) =\sum_{i}{\bf D}_{i}^{+}\left( {\bf r}%
\right)   \label{jellium1}
\end{equation}
where
\[
{\bf D}_{i}^{+}\left( {\bf r}\right) = \left\lbrace
\begin{array}{lcl}
\displaystyle \frac{e}{4\pi R_{s}^{3}}\left( {\bf r-R%
}_{i}\right) & ;& \mbox{if} \; \;  \left| {\bf r-R}_{i}\right| <R_{s} \\
\displaystyle \frac{e}{4\pi }\frac{{\bf r-R}_{i}}{%
\left| {\bf r-R}_{i}\right| ^{3}} &;& \mbox{otherwise}
\end{array}
\right.
\]
so that
\begin{eqnarray}
\frac{1}{8\pi \varepsilon _{s}}\int {\bf D}_{i}^{+}\cdot {\bf D}_{i}^{+} &=&%
\frac{3e^{2}}{5\varepsilon _{s}R_{s}}  \nonumber \\
\frac{1}{8\pi \varepsilon _{s}}\sum_{i\neq j}\int {\bf D}_{i}^{+}\cdot {\bf D%
}_{j}^{+} &=&\frac{1}{2}\sum_{i\neq j}\frac{e^{2}}{\varepsilon _{s}\left|
{\bf R}_{i}-{\bf R}_{j}\right| }  \label{energy jellium}
\end{eqnarray}

We next assume that the electrons are localized around the sites ${\bf R}%
_{i} $ of the same Bravais lattice, and  we introduce the small
displacements ${\bf u}_{i}={\bf r}_{i}-{\bf R}_{i}$, so that
\begin{eqnarray}
\frac{1}{4\pi \varepsilon _{s}}\int_{(i\neq j)}{\bf D}_{i}^{+}\cdot {\bf D}%
_{j}^{-} &=&-\frac{ e^{2}}{\varepsilon _{s}\left| {\bf R}_{i}-{\bf R}%
_{j}-{\bf u}_{j}\right| }  \nonumber \\
\frac{1}{4\pi \varepsilon _{s}}\int {\bf D}_{i}^{+}\cdot {\bf D}_{i}^{-} &=&-%
\frac{3e^2}{2\varepsilon _{s}R_{s}}+\frac{m^{*}}{2}\frac{\omega _{W}^{2}}{%
\varepsilon _{s}}u_{i}^{2}  \label{energy el-jel}
\end{eqnarray}
where we have defined $\omega _{W}^{2}=e^2/m^*R_s^3=\omega _{p}^{2}/3$,
$\omega_{p}$ being the electron plasma frequency.
Both  $S_{e}$ and $S_{e-ph}$ can be decomposed into diagonal ($i=j$) and
off-diagonal ($i\neq j$) terms, i.e.:
\begin{eqnarray}
\lefteqn{S_{e}  = -\frac{e^{2}}{2\varepsilon _{\infty }}\sum_{i\neq
j}\int_{0}^{\beta }\frac{%
1}{\left| {\bf R}_{i}-{\bf R}_{j}+{\bf u}_{i}(\tau )-{\bf u}_{i}(\tau
)\right| }d\tau } \nonumber \\
&&-\sum_{i}\int_{0}^{\beta
}\left[ \frac{9e^{2}}{10\varepsilon _{s}R_{s}}+\frac{m^{*}}{2}\dot{u}%
_{i}^{2}\left( \tau \right) +\frac{m^{*}}{2}\frac{\omega _{W}^{2}}{%
\varepsilon _{s}}u_{i}^{2}\left( \tau \right) \right] d\tau \nonumber
\end{eqnarray}
and
\begin{eqnarray}
&&S_{e-ph} =\sum_{i}\frac{\omega
_{LO}e^{2}}{4{\tilde{\varepsilon}}}{{\int_{0}^{\beta }{}}{\int_{0}^{\beta }{%
\frac{G_{\omega _{LO}}(\beta ,\tau -\sigma )}{|{\bf u}_{i}(\tau )-{\bf u}%
_{i}(\sigma )|}}d\tau d\sigma }}  \nonumber \\
&&+\sum_{i\neq j}\frac{\omega _{LO}e^{2}}{4{\tilde{\varepsilon}}}{{%
\int_{0}^{\beta }{}}{\int_{0}^{\beta }{\frac{G_{\omega _{LO}}(\beta ,\tau
-\sigma )}{|{\bf R}_{i}-{\bf R}_{j}+{\bf u}_{i}(\tau )-{\bf u}_{i}(\sigma )|%
}}d\tau d\sigma }}  \nonumber
\end{eqnarray}

Our next approximation
consists in
expanding the action up to second order in the $\left\{ {\bf
u}_{i}\right\}$,  thus ignoring all the
anharmonic and higher order contributions. After some elementary algebra,
one gets:
\begin{equation}
S\left( \left\{ {\bf u}_{i}\right\} \right) =\sum_{i}S_{i}+\frac{1}{2}\sum_{i%
\neq j}S_{ij}  \label{Sfinal}
\end{equation}
with
\begin{eqnarray}
S_{i} &=&-\beta \frac{9e^{2}}{10\varepsilon _{s}R_{s}}-\int_{0}^{\beta
} \left[ \frac{m^{*}}{2}\dot{u}_{i}^{2}\left( \tau \right) +\frac{m^*}{2}%
\frac{\omega _{W}^{2}}{\varepsilon _{s}}u_{i}^{2}\left( \tau \right)
\right] d\tau  \nonumber \\
&&+\frac{\omega _{LO}e^{2}}{4{\tilde{\varepsilon}}}{\int_{0}^{\beta } \!
{\int_{0}^{\beta }{\frac{G_{\omega _{LO}}(\beta ,\tau -\sigma )}{|{\bf u}%
_{i}(\tau )-{\bf u}_{i}(\sigma )|}d\tau d\sigma }}}  \label{S-diag}
\end{eqnarray}
and
\begin{eqnarray}
& &{S_{ij} =-\frac{e^{2}}{{\varepsilon }_{\infty }}\sum_{\alpha \gamma
}\int_{0}^{\beta }\Lambda _{ij}^{\alpha \gamma }u_{i}^{\alpha }\left( \tau
\right) u_{j}^{\gamma }\left( \tau \right)  d\tau} \label{S-off-diag} \\
& &+\frac{\omega _{LO}e^{2}}{2{\tilde{\varepsilon}}}\sum_{\alpha \gamma
}\int_{0}^{\beta } \! \int_{0}^{\beta }\Lambda _{ij}^{\alpha \gamma }G_{\omega
_{LO}}(\beta ,\tau -\sigma )u_{i}^{\alpha }\left( \tau \right)
u_{j}^{\gamma }\left( \sigma \right) {d\tau d\sigma }  \nonumber
\end{eqnarray}
where the indices $\alpha ,\gamma =(x,y,z)$ denote the cartesian
coordinates. We have also defined the dipolar matrix elements:
\begin{equation}
\Lambda _{ij}^{\alpha \gamma }=\frac{\delta _{\alpha \gamma
}R_{ij}^{2}-3R_{ij}^{\alpha }R_{ij}^{\gamma }}{R_{ij}^{5}}
\end{equation}
Expressions (\ref{Sfinal})-(\ref{S-off-diag}) constitute our basic
model for the polaron crystal. Since we focus on the insulating state at
low density, the action $S$ can be treated semi-classically, i.e.
neglecting the exchange between the fermions and quantum statistical
effects.

\subsection{Comparison with the electron crystal}

Our model (\ref{Sfinal}) was obtained by making use of the
following
approximations: (i) we consider a particular structure for the jellium,
(ii) we expand the off-diagonal part of the action up to second order in
the displacements
$\mathbf{u}_i$, and (iii) we neglect the exchange between different electrons.
In order to analyse in more detail the consequences of these assumptions,
we can compare our results with the previous treatments of the
electron crystal. Carr \cite{carr} demonstrated that the
energy per electron in a WC with bcc symetry can be expressed
as an expansion in powers of $R_{s}^{-1/2}$. Introducing the dimensionless
density parameter $r_s=R_s/a_{0}$ ($a_{0}=\hbar^2/m_ee^2 \approx 0.53$\AA \
is the Bohr radius)  and taking the Rydberg as the energy unit, the
result reads
\begin{equation}
E=\frac{-1.792}{r_{s}}+\frac{2.66}{r_{s}^{3/2}}+\frac{b}{r_{s}^{2}}%
+O(r_{s}^{5/2})+O\left( e^{-r_{s}^{1/2}}\right)  \label{E-Wigner}
\end{equation}
with $b<1$.
Basically, our three approximations above consist in neglecting all the
terms in powers of $r_{s}$ lower than $-3/2$ in (\ref{E-Wigner}).  In fact,
the first assumption slightly overestimates the Madelung energy, whose
numerical value is lowered from $-1.792/r_{s}$ to $-1.8/r_{s}$.  The second
hypothesis consists in neglecting the anharmonic ($b/r_{s}^{2}$) and
higher order terms.  These terms can become important close to the melting
point, and can change its numerical determination, but do not affect
qualitatively the physics of the problem.  Our last approximation is to
neglect all the terms proportional to the overlap between the wavefunctions
of different localized electrons.  The corresponding energy terms in
(\ref{E-Wigner}), i.e. the exchange terms, fall off exponentially with
$r_{s}^{1/2}$.  As pointed out
by Carr, they are negligible up to $r_{s}\approx 10$.  In the polaron
crystal, where localization is more efficient due to the e-ph coupling, it
is quite reasonable to neglect them in the density range of interest (these
terms are responsible for the magnetic properties of the crystal, that we
do not consider here).

\section{The mean-field Wigner approximation: solution by the
Feynman method}
\label{sec:Wigner}

In the following, we will restrain ourselves to the diagonal part of the action
$S=\sum_{i}S_{i}$, neglecting the
dipolar terms $ S_{ij} $.
This is generally called the Wigner approximation.
It is sensitive since, using Gauss' theorem, it can be
easily shown that the total mean electric field of a
particular sphere (averaged in time), including the jellium, the electron,
and the polarization
field, vanishes for $|\mathbf{r}_i-\mathbf{R}_i|>R_s$
\cite{quem1}.
In other words, two different Wigner spheres can only
interact through the deviations of $\mathbf{r}_i$ from equilibrium,
and that goes beyond the mean-field approximation. Each sphere
is then \textit{on the average} an independent entity, and the effect of the
other particles is already included in the harmonic potential of eq.
(\ref{energy el-jel}).
At this stage, all the electrons
localized in their jellium spheres are equivalent and uncoupled, so that
the system is fully described by any of the single-particle actions
(\ref{S-diag}) --- i.e. we can take $S=S_i$. As was stated in the introduction,
the off-diagonal elements
$S_{ij}$ will be treated in a further publication.

\begin{figure}[ht]
\centerline{\resizebox{8cm}{!}{\includegraphics{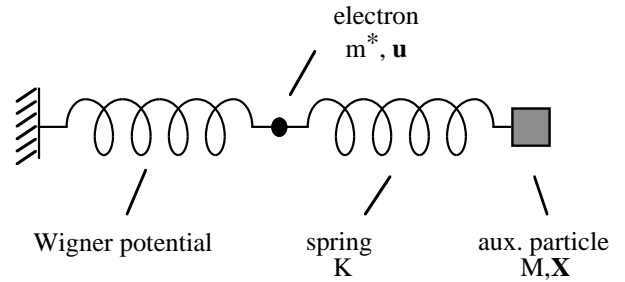}}}
\caption{The classical model corresponding to the action (\ref{qaction}): the
polarization cloud is replaced by a fictitious particle of mass $M$.
\label{fig-feyn-model}}
\end{figure}

The retarded interaction in (\ref{S-diag}), which is responsible
for polaron formation, cannot be integrated exactly.
%
Following Feynman
\cite{feynman}, we introduce a quadratic trial action of the form
\begin{eqnarray}
S_{0} &=&-\frac{m^{*}}{2}\int \dot{ u}^{2} d\tau -\frac{m^{*}}{2}\frac{%
\omega _{W}^{2}}{\varepsilon _{s}}\int  u^{2} d\tau  \nonumber \\
&-&\frac{Kw}{8}\int \int {|{\bf u}(t)-{\bf u}(s)|}%
^{2}G_{w}(\beta ,\tau -\sigma )d\tau \,d\sigma ,  \label{qaction}
\end{eqnarray}
whose parameters $K$ and $w$ will be determined variationally
(here $\mathbf{u}_i$ has been replaced by the coordinate $\mathbf{u}$ of the
only electron present). It can be
seen that (\ref{qaction}) corresponds to the following two-body Lagrangian
\begin{equation}
L_{0}=\frac{m^{*}}{2}{\dot{u}}^{2}+\frac{M}{2}{\dot{X}}^{2}-\frac{K
}{2}\left( {\bf u-X}\right) ^{2}-\frac{m^{*}}{2}\frac{\omega _{W}^{2}}{%
\varepsilon _{s}}{u}^{2}.  \label{L_0}
\end{equation}
after the integration of $\mathbf{X}$ has been carried out.
This model describes
an electron  subject to an external harmonic potential and bound
with a spring to a fictitious particle of mass $M$ and coordinate ${\bf
X}$. The latter represents the polarization
cloud associated to the electron (see Fig.\ref{fig-feyn-model}).
The constant $K$, which measures the strength of the retarded attraction,
can be written as $K=Mw^2$.

 The Lagrangian (\ref{L_0}) can be diagonalized into two harmonic modes $%
{\bf u}_{1}$ and ${\bf u}_{2}$, whose
 eigenfrequencies $\alpha _{1}$ and
$\alpha _{2}$ are  given by:
\begin{equation}
\alpha _{1,2}^{2}=\frac{v^{2}+\omega _{W}^{2}/\varepsilon _{s}}{2}\mp \frac{%
\sqrt{\left( v^{2}+\omega _{W}^{2}/\varepsilon _{s}\right) ^{2}-4w^{2}\omega
_{W}^{2}/\varepsilon _{s}}}{2}  \label{eigenfrequencies}
\end{equation}
We have introduced the shortcut notation $v=\left( K/\mu \right) ^{1/2}$
for the characteristic frequency of the two-body model
($\mu =m^{*}M/\left( m^{*}+M\right) $ is
the reduced mass). The two eigenfrequencies obey the following
relations:
\begin{eqnarray}
\alpha _{1}^{2}+\alpha _{2}^{2} &=&v^{2}+\omega _{W}^{2}/\varepsilon _{s}
\nonumber \\
\alpha _{1}\alpha _{2} &=&w\omega _{W}/\sqrt{\varepsilon _{s}}
\label{relation(alpha)} \\
\alpha _{1} &\leq &w\leq \alpha _{2}.  \nonumber
\end{eqnarray}
Finally, the mass $M$ of the auxiliary particle can be written as:
\begin{equation}
\frac{M}{m^{*}}=\left( 1-\frac{\alpha _{1}^{2}}{w^{2}}\right) \left( \frac{
\alpha _{2}^{2}}{w^{2}}-1 \right) ,  \label{relation(mass)}
\end{equation}
and the polaron mass is defined as:
\begin{equation}
        M_{P}=m^{*}+M.
        \label{pol-mass}
\end{equation}
If we introduce the quantities
\[
A_{1}=\frac{w^{2}-\alpha _{1}^{2}}{\alpha _{2}^{2}-\alpha _{1}^{2}},A_{2}=%
\frac{\alpha _{2}^{2}-w^{2}}{\alpha _{2}^{2}-\alpha _{1}^{2}},
\]
the relation between the real coordinates and the normal modes
is simply:
\begin{eqnarray}
\mathbf{u} &=&A_{1}\mathbf{u}_{1}+A_{2}\mathbf{u}_{2}  \nonumber \\
\mathbf{X} &=&\frac{w^{2}}{\alpha _{2}^{2}-\alpha _{1}^{2}}
\left( \mathbf{u}_{1}-\mathbf{u}_{2}\right) . \label{coord}
\end{eqnarray}
and the diagonalized Lagrangian can be expressed in canonical form:
\begin{equation}
L_{0}=\frac{1}{2}A_{1}{ \dot{u}}_{1}^{2}+\frac{1}{2}A_{2}{ \dot{u}}%
_{2}^{2}-\frac{1}{2}A_{1}\alpha _{1}^{2}{ u}_{1}^{2}-\frac{1}{2}%
A_{2}\alpha _{2}^{2}{ u}_{2}^{2}. \label{Ldiagx}
\end{equation}

\subsection{The low density regime}
\label{sec-lowdens}
\begin{figure}[ht]
\resizebox{8cm}{!}{\includegraphics{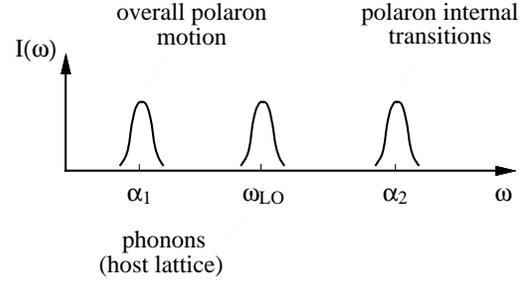}}
\caption{Schematic absorption spectrum of the polaron crystal within
Wigner's approximation. The two
energy scales $\alpha_1$ and $\alpha_2$ are well separated, and
the (host) phonon frequency lies in between. \label{fig-energyscale}}
\end{figure}

At very low densities ($r_{s}\rightarrow \infty $, $\omega _{W}
\approx 0$),
the polarons are so far apart that their properties are
unchanged from the single polaron case. In other words,
the localizing potential $V(\mathbf{ u}
)=m^{*}\omega _{W}^{2}u^{2}/2\varepsilon_s$ acting on the electron is a
perturbation with respect to the polaron energy, and
the eigenfrequencies (\ref
{eigenfrequencies}) can be expanded for small $\omega
_{W}^{2}/\varepsilon _{s}$, which gives:
\begin{eqnarray}
\alpha _{1} &\rightarrow &\sqrt{\frac{m^{*}}{M_{P}}}\cdot
\frac{\omega _{W}}{\sqrt{\varepsilon _{s}}} \equiv \omega_{ext} \nonumber \\
\alpha _{2} &\rightarrow &v.  \label{limit(frequence)}
\end{eqnarray}
($\omega_{ext}$ is defined as the frequency of vibration of a particle of
mass $M_P$ in a harmonic potential whose spring constant is
$m^*\omega_W^2/\varepsilon_s$, while $v$ is the internal frequency of an
isolated
polaron \cite{remark-notations}).
The corresponding eigenmodes are:
\begin{eqnarray}
{\bf u}_{1} &\rightarrow &{\bf R=}\frac{m^{*}{\bf u}+M{\bf X}}{M_{P}}
\nonumber \\
{\bf u}_{2} &\rightarrow &{\bf r}={\bf u}-{\bf X.}  \label{limit(mode)}
\end{eqnarray}
These two   independent modes correspond respectively to the vibration
of the polaron center-of-mass, and to the vibration of the
electron inside the polarization potential well. We shall call them
respectively the {\it external} and the {\it internal} degree of
freedom. It can be seen from  expressions (\ref{limit(frequence)})  that
in the dilute regime,  the external and internal energy scales are well
separated, and
the polaron can follow the vibrations imposed by the
external potential as if it was a rigid particle.
Basically, $\alpha_1$ and $\alpha_2$ give the excitation spectrum of the
polaron crystal in the framework of the Wigner approximation, as is
sketched in Fig.\ref{fig-energyscale}.
(note that the electron can also be excited towards a free state outside
the polaron potential; the corresponding frequency, which is higher
than $\alpha _{2}$, is not shown here).

\subsection{The crossover regime}
\label{sec:crossover}

As was noted in the previous section, the localizing Wigner
potential forces the polaron center-of-mass to vibrate with a frequency
$\omega _{ext}$, which increases with
increasing density. However, the polaron is a composite particle, and its
velocity is physically limited
by the characteristic frequency of the phonon cloud:
when $\omega _{ext}$ approaches $\omega _{LO}$, the phonons
are too slow to ''follow'' the overall motion, and only the electronic
degree of freedom
oscillates.
In that case,  both $\alpha _{1}$ and $\alpha _{2}$ can
deviate significantly from their asymptotic values (\ref{limit(frequence)}).
If the density is further increased, the
excess kinetic energy is transferred to the internal
degree of freedom and the concept of a rigid polaron vibrating in an
external potential breaks down.
The crossover density is roughly given by the condition:
\begin{equation}
\omega_{ext} \approx \omega _{LO}  \label{crossovercondition}
\end{equation}
that we express using (\ref{limit(frequence)}) as:
\[
\left( \frac{m^{*}}{m_{e}}\right) \frac{r_{s}}{\varepsilon _{\infty }}%
\approx \left[ \left( \frac{m^{*}}{M_{P}}\right) \frac{4\eta }{\left( 1-\eta
\right) ^{4}}\alpha ^{4}\right] ^{1/3},
\]
where $\eta =\varepsilon _{\infty }/\varepsilon _{s}$.
In the limit of strong e-ph coupling  ($\alpha \gtrsim 6$), the
 mass of an isolated polaron is well approximated by $M_{P}/m^{*}\approx %
0.02\alpha ^{4}$ \cite{feynman,schultz}, so that eq.
(\ref{crossovercondition}) can be
expressed as:
\begin{equation}
\left( \frac{m^{*}}{m_{e}}\right) \frac{r_{s}}{\varepsilon _{\infty }}%
\approx \left[ 200\frac{\eta }{\left( 1-\eta \right) ^{4}}\right] ^{1/3}
\end{equation}
and the crossover density becomes independent
on the e-ph coupling. As an example,
if we take $m^*=m_{e}$, $\varepsilon _{s}= 30$ and $\varepsilon _{%
\infty }= 5$, the crossover region corresponds to
$r_{s}\approx 20$.
In the opposite limit ($\alpha \rightarrow 0$), the phonon frequency
tends to infinity, so that $\omega_{ext}$ never reaches the value
$\omega_{LO}$.
In that case, the internal structure of the polarons can
safely be neglected, and the polaron crystal tends to an ordinary electron
crystal.

\subsection{Results of the variational procedure}

\begin{figure}[ht]
\resizebox{8cm}{!}{\includegraphics{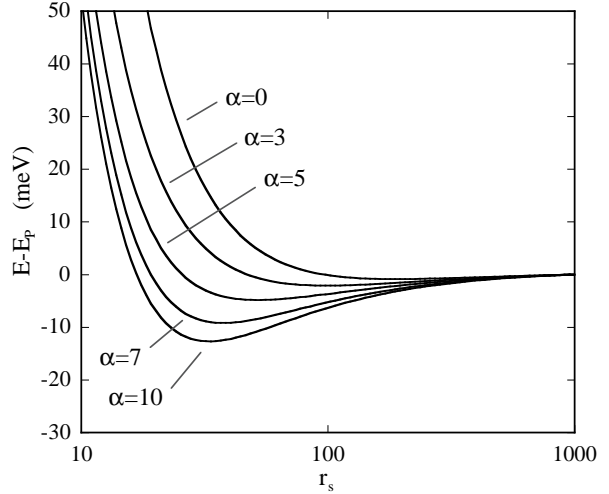}}
        \caption{The energy per polaron in the crystallized state versus $r_s$
        (the single polaron value $E_P$,  corresponding to the limit $r_s=\infty$, has
        been subtracted out). From
        top to bottom $\alpha=0$ (electron crystal), $\alpha=3,5,7,10$. The
        energy unit is $meV$. In this and the following figures, the parameters
        of the host lattice are $\varepsilon_s=30$, $\varepsilon_\infty=5$,
        $m^*=m_e$.}
        \label{fig-energy}
\end{figure}

As in the case of the single polaron problem,  Feynman's
variational method gives, for any $\alpha$, the best analytical
upper bound for the free energy $F$.
This is obtained by minimizing the following expression:
\begin{equation}
F=F_{0}-\frac{1}{\beta }\langle
S-S_{0}\rangle   \label{Var-F}
\end{equation}
where the term $F_{0}$ is defined as
\begin{equation}
e^{-\beta F_{0}}=\int {\cal D}(path)\,e^{S_{0}}.  \label{path_energy}
\end{equation}
and $\langle \cdots \rangle $ stands for $\int {\cal D}(path)\,(\ldots)
e^{S_{0}}/\int {\cal D}(path)\,e^{S_{0}}$ (path integrations must be
carried out on closed trajectories).
The calculation of the functional (\ref{Var-F}) to be minimized follows the
lines of Feynman's work (details are presented in the Appendix).  The
ground-state energy $E$ at zero temperature can be obtained by taking the
limit $\beta\rightarrow \infty$ in the preceding expressions.  The result
is:
\begin{eqnarray}
E &=&C_0 +%
\frac{3}{2}(\alpha _{1}+\alpha _{2}-w)
-\frac{3(w^{2}-\alpha
_{1}^{2})(\alpha _{2}^{2}-w^{2})}{4(\alpha _{1}+\alpha _{2})w^{2}}   \nonumber
\\ &-&\frac{\alpha }{\sqrt{\pi }}\int_0^\infty
\frac{e^{-t}\,dt}{\sqrt{\frac{A_{1}}{%
\alpha _{1}}\left( 1-e^{-\alpha _{1}t}\right) +\frac{A_{2}}{\alpha _{2}}%
\left( 1-e^{-\alpha _{2}t}\right) }}  \label{E_var}
\end{eqnarray}
where  all frequencies are expressed in units of
$\omega_{LO}$, and
$C_0=-(9\tilde{\varepsilon}^2\alpha^2/5\varepsilon_sr_s)(m_e/m^*)$.
The last integral is non elementary and must be calculated numerically. $%
\alpha _{1,2}$ are taken as the two independent variational parameters, and
 $w$ is related to $\alpha _{1,2}$ through (\ref
 {relation(alpha)}).
The expression (\ref{E_var}), after minimization,
represents the energy per particle of the PWC in the Wigner
approximation. The result is sketched in Fig.\ref{fig-energy} for
different values of $\alpha$.
It is interesting to note that, upon increasing $\alpha$, the evolution
from  a WC of electrons towards a PWC is gradual. In fact, all the curves
in Fig.\ref{fig-energy} have
the same behaviour, and the only difference comes from the fact that the
minimum of $E$ becomes more pronounced and shifts to higher densities.
This is not surprising, since it is known  that the
formation of large polarons is not a real phase transition, but rather a
continuous crossover \cite{gerlach}.

\subsubsection{Eigenfrequencies}

\begin{figure}[ht]
\resizebox{8cm}{!}{\includegraphics{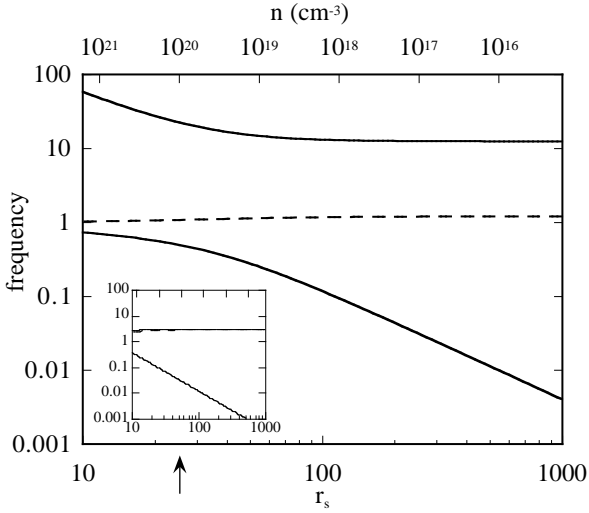}}
        \caption{The characteristic frequencies of the two-body model
        (\ref{L_0}), in units of $\omega_{LO}$: $\alpha_2$
        (upper curve), $w$ (dashed curve), $\alpha_1$ (bottom curve). The e-ph
        coupling is $\alpha=10$. The arrow marks the crossover region (see sec.
        \ref{sec:crossover}). Inset: same as main plot, with $\alpha=1$.
        }
        \label{fig-frequency}
\end{figure}

The three frequencies $w$, $\alpha_1$ and $\alpha _2$
are illustrated in Fig.\ref{fig-frequency} as a function of the density,
for two different values of $\alpha$.
As noted above, at weak e-ph coupling, the phonon cloud is fast and it
can easily follow the polaron motion at any density. The external
frequency $\alpha_1$ is proportional to $\omega_W$, while $\alpha_2$,
which tends to $w$ in this limit, is almost density independent (see inset).
The two
frequencies are always well separated, and the polaron can vibrate as if
it was a rigid particle.

On the other hand, at intermediate and strong $\alpha$, the internal and
external degrees of freedom are fully decoupled only at low density
(see main plot). When we reach the crossover region (indicated by an arrow),
the external frequency "saturates" around $\omega_{LO}$: the phonon
cloud  cannot follow the vibrations imposed by
the Wigner potential, and the polaron center-of-mass is virtually frozen.
Upon increasing the density, the excess kinetic energy is thus
transferred to the internal degree of freedom, and $\alpha_2$ becomes
proportional to $\omega_W$.

%
%

\subsubsection{Polaron mass}

\begin{figure}[ht]
\resizebox{8cm}{!}{\includegraphics{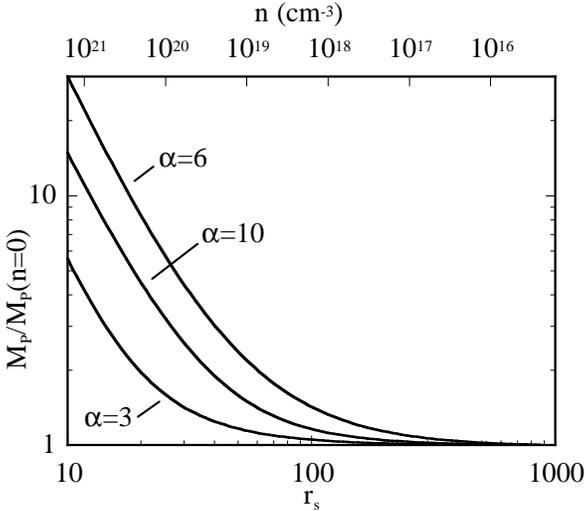}}
        \caption{The polaron mass $M_P$, normalized by the single polaron value,
        for $\alpha=3,6,10$.}
        \label{fig-mass}
\end{figure}

In Fig.\ref{fig-mass} we illustrate the behaviour of $M_P$ as a function of
$r_s$, for different values of $\alpha $.
It can be
seen that $M_P$ always increases with the density when one approaches the
crossover regime. A similar effect is well-known for an isolated polaron
moving at a finite velocity $\mathrm{v}$ \cite{kartheuser}. If
the quasi-particle kinetic energy $M_{P}\mathrm{v}^{2}/2$ reaches the
phonon energy,
the effective
polaron mass increases because  phonons can be emitted from
the polarization cloud in an incoherent way (the equivalent of Cherenkov
effect in transparent media). If we consider
a polaron in a Wigner sphere, whose average velocity is $\mathrm{v}\approx
\omega_{ext}\sqrt{\hbar /M_{P}\omega_{ext}}$, it is easy to
verify that  the condition $M_{P}\mathrm{v}^{2}/2\approx \hbar \omega
_{LO}$ gives
exactly the crossover condition (\ref{crossovercondition}).

\section{The insulator-to-metal transition}

The melting of the ordinary electron crystal at zero temperature has been
extensively studied in the literature, and there exist many different
criteria which allow to evaluate the critical density for the transition
(see Care \cite{care} for a review).  A particularly simple approach was
applied by Nozi\`{e}res and Pines \cite{nozieres}, which is based on the
Lindemann criterion.  This phenomenological criterion,
generally used to describe the thermal melting of the atomic lattices,
states that the crystal is unstable when the spatial
fluctuations of each particle around its equilibrium position
exceed some fraction of
the inter-particle distance.  Such criterion will be generalized here to the
case of a PWC, and will be used at both zero and
finite temperature in order to describe qualitatively the
insulator-to-metal transition.


\subsection{Lindemann Criterion}
\label{sec:LindT0}
Calling $\mathbf{u}=(x,y,z)$ the coordinate of a localized electron in an
ordinary Wigner
crystal, the Lindemann criterion states that the crystal melts when
\[
\langle \delta u^{2}\rangle ^{1/2}/R_{s}>\delta
\]
where $ \delta $ is a phenomenological constant usually taken as $\delta
\approx
0.25 $. In the Wigner approximation,  the average zero-point displacement in
each space direction (say $x$) is given at zero temperature  by:
\begin{equation} \left\langle \delta
x^{2}\right\rangle ^{1/2}=\left( \frac{\hbar }{ 2m_e\omega _{W}}\right)
^{1/2}.
\end{equation}
If we consider three dispersionless modes of
equal frequency $\omega _{W}=\omega _{p}/\sqrt{3}$,
as  is the case if we  neglect
dipole-dipole interactions, we get a critical density parameter
$r_{c}^{W}= 64$,
which is close to the
Monte-Carlo result  ($r_{c}^{W}= 100\pm 20$ in 3D, \cite{MC3D}).
Of course, this criterion is only qualitative and the resulting
melting density is very sensitive to the choice of $\delta$, and moreover
 it does not give any precise information about the nature of
the ground-state beyond the melting point.
However, especially in the polaronic case, we shall see that it gives
a good insight in the physics of the problem.

The treatment that we have described in the previous sections,
which reduces the many-polaron problem to the problem of a single polaron
in an effective potential, gives reliable results for the insulating state
 at any value of the e-ph coupling.
It was shown in  section \ref{sec:Wigner} that  the system passes
continuously from an ordinary Wigner lattice at  weak $\alpha$
 to a lattice of polarons at strong
$\alpha$, with no symmetry breaking between the two limits. However,
we will show here that a definite distinction between the two cases comes
from the
melting mechanisms.
Indeed, taking into account the
composite nature of the polarons, we can introduce
 two different Lindemann criteria, whether we analyse  the
fluctuation of the polaron as a whole, or rather  the fluctuation of
the electron with respect to the polaritazion field.
Using the
definitions (\ref{limit(mode)}) of the center-of-mass ${\bf R}$ and of the
relative coordinate ${\bf r}$, we can write the following conditions:
\begin{eqnarray}
(i)&  & \langle \delta R^{2}\rangle ^{1/2}/R_{s}>0.25 \nonumber \\
(ii)& & \langle \delta r^{2}\rangle ^{1/2}/R_{s}>0.25 \nonumber
\end{eqnarray}
The former gives the critical
value $r_{c}^{\left( i\right) }$ for the melting towards a polaron liquid.
The latter gives the critical value $r_{c}^{\left( ii\right) }$ for
 polaron dissociation.
 The calculation of $\langle \delta R^{2}\rangle$ and $\langle \delta
r^{2}\rangle$
in terms of the
variational parameters is presented in the appendix. The result at $T=0$
is (all frequencies are expressed in units of $\omega_{LO}$):
\begin{eqnarray}
\langle \delta R^{2}\rangle ^{1/2}/R_{s} &=&\left( \frac{\omega
_{W}^{2}}{4\alpha }
\right) ^{1/3}\frac{m^*}{%
M_P}\sqrt{\frac{A_{1}}{\alpha _{1}}\left( \frac{\alpha _{2}}{w}\right) ^{4}+%
\frac{A_{2}}{\alpha _{2}}\left( \frac{\alpha _{1}}{w}\right) ^{4}}  \nonumber
\\ \nonumber
\langle \delta r^{2}\rangle ^{1/2}/R_{s} &=&\left( \frac{\omega
_{W}^{2}}{4\alpha }\right) ^{1/3}\frac{1}{%
\alpha _{2}^{2}-\alpha _{1}^{2}}\sqrt{\frac{\alpha _{1}^{2}}{A_{1}}+\frac{%
\alpha _{2}^{2}}{A_{2}}}.
\end{eqnarray}
When the e-ph coupling vanishes ($M_P\rightarrow m^{*}$ and
$\mathbf{R}\rightarrow \mathbf{u}$),
eq. $(i)$  reduces to the ordinary Lindemann
criterion for the WC, and the transition is due to
fluctuations of the  (weakly renormalized) electrons.
On the other hand, the fluctuation of the internal coordinate gives a
measure of the polaron radius \cite{schultz}, so that criterion
$(ii)$ can be written as $R_{P}/R_{s}=const$, which closely resembles
eq. (\ref{Mott-gen}) above. As a matter of fact, we can consider
the second criterion  as a generalization to finite phonon frequencies
of the dissociation argument given in the introduction (which we have shown to
be valid in the limit
$\omega_{LO}\rightarrow 0$, when the potential acting on the electron is
completely static).

\subsection{Transition at T=0}


 We are now in a position to analyse the IMT
for any values of $\alpha $.
\begin{figure}[ht]
\resizebox{9cm}{!}{\includegraphics{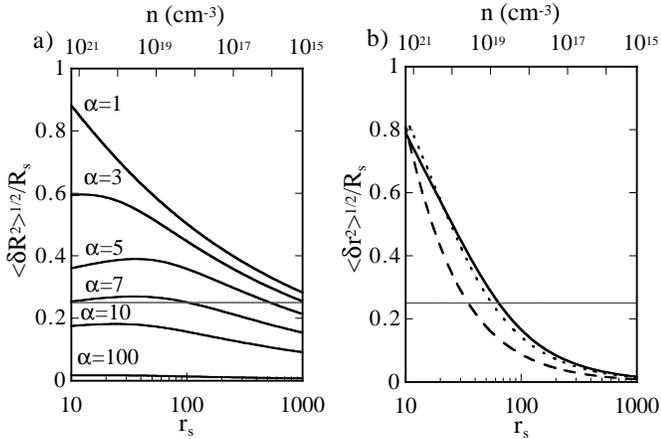}}
        \caption{The ratios which enter in the Lindemann criteria $(i)$ and
        $(ii)$: a) $\langle \delta R^{2}\rangle ^{1/2}/R_{s}$ for
        $\alpha=1,3,5,7,10,100$; b) $\langle \delta r^{2}\rangle ^{1/2}/R_{s}$
        for $\alpha=1$ (dashed line), $\alpha=3$ (dotted line), $\alpha=10$
        (full line). The horizontal line corresponds to the phenomenological
        melting value $\delta=0.25$.}
        \label{fig-radius}
\end{figure}
Fig.\ref{fig-radius} shows the calculated ratios which enter
in the conditions $(i)$ and $(ii)$, as functions of $r_{s}$, for
different values of $\alpha$. One observes
that for a fixed $\alpha $, while $\langle \delta r^{2}\rangle ^{1/2}/R_{s}$
always increases with increasing density (all the curves in
Fig.\ref{fig-radius}.b have roughly the
same behaviour), the center-of-mass
fluctuation $\langle \delta R^{2}\rangle ^{1/2}/R_{s}$ becomes bell
shaped for sufficiently large $\alpha$ (it changes its
slope exactly where, according to the argument of section \ref{sec:crossover},
the internal structure
of the polaron becomes important --- see Fig.\ref{fig-radius}.a).
In addition, the maximum value of $\langle \delta R^{2}\rangle
^{1/2}/R_{s}$ decreases
with increasing $\alpha $. Remembering that the transition occurs when one
of the fluctuation ratios reaches the value $0.25$, we see that there exists
a certain critical value $\alpha ^{*}$ above which the crystal melting
\textit{\`a la Wigner} is prevented, and the driving mechanism switches to the
polaron dissociation, typical of the static limit.
An important consequence of this result is that, for $\alpha
>\alpha ^{*}$, the polarons do not survive beyond the IMT. In other
words, in the strong e-ph coupling limit,   a
liquid state of large polarons cannot exist at zero temperature. With the
parameters
of Fig. \ref{fig-radius}, we find $\alpha ^{*}\simeq 7.5$.


\bigskip

\begin{figure}[ht]
 \resizebox{8cm}{!}{\includegraphics{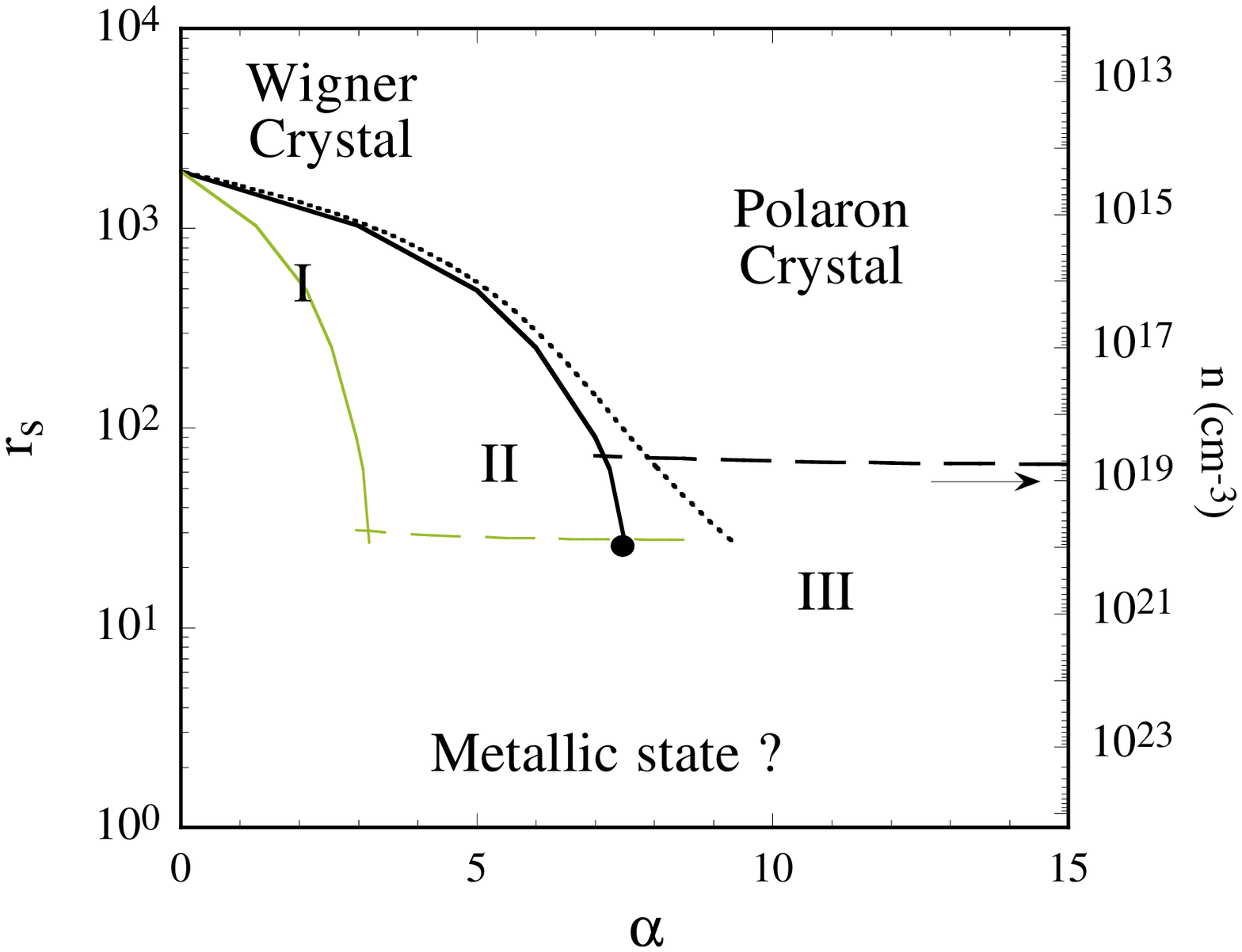}}
        \caption{Zero temperature phase diagram as deduced from criteria $(i)$
        and $(ii)$: $r_c^{(i)}$ for crystal melting  (full line) and $r_c^{(ii)}$
        for polaron dissociation in 3D (dashed line).
        The corresponding quantities for an anisotropic polaron crystal (see section
        \ref{sec:anisotropic}) are drawn in grey.
        The dotted line corresponds to
        the low density approximation (\ref{Rcrit}). The melting driven by the
        fluctuations of the localized particles (criterion $(i)$) is impossible
        beyond $\alpha^*$ (full circle). The critical density given by eq.
        (\ref{Mott-gen}) is indicated by an arrow. For regions I, II and III,
        see text.}
        \label{fig-phasediag}
\end{figure}
Fig.\ref{fig-phasediag} illustrates the approximate phase diagram
obtained from the two Lindemann criteria at $T=0$.
The full line corresponds to criterion $(i)$. We see that $r_c^{(i)}$ decreases
with increasing $\alpha$ (the corresponding
critical density increases). This can be easily
understood in terms of the asymptotic behaviour of section
\ref{sec-lowdens}.
As we have stated above, at low density,
we can neglect the composite nature of the polaron and its mean square
displacement in an external potential of frequency $\omega _{ext}$ is given
by:
\begin{equation}
\langle \delta R^{2}\rangle ^{1/2}\simeq\sqrt{\frac{\hbar }{2M_{P}\omega
_{ext}}},
\label{R_low}
\end{equation}
which fits very well the low density part of $\langle \delta R^{2}\rangle
^{1/2}/R_{s}$ (right side of Fig. \ref{fig-radius}.(a)).
In the weak coupling regime (in practice for all $\alpha <\alpha ^{*}$), we can
use this formula to calculate the critical $r_s$ for
crystal melting. The
result can be expressed in terms of the critical parameter $r_{c}^{W}$ of
the ordinary electron
crystal as:
\begin{equation}
r_{c}^{\left( i\right) }\simeq r_{c}^{W}\varepsilon _{s}\frac{m_e}{M_{P}}.
\label{Rcrit}
\end{equation}
The competition between the kinetic and potential energy
is clearly visible in eq. (\ref{Rcrit}): if the mass of the carriers is
strong enough,
despite the static
screening of the charges which reduces the Coulomb interactions, the
crystallization is favored with respect to the ordinary WC of
electrons.
The melting curve corresponding to this approximation
is shown in Fig.\ref{fig-phasediag} (dotted line: we see
that (\ref{Rcrit}) slightly
underestimates the critical density, especially for $\alpha$  close to
$\alpha^*$).  Since the polarons do not dissociate at the transition, we
expect the metallic state beyond $n_c^{(i)}$
to be either a renormalized electron liquid (region I) or a degenerate polaron
liquid (region II).

For $\alpha>\alpha^*$, the IMT is described by the
dissociation mechanism $(ii)$, and the critical density
 is almost constant
(in a nutshell, the crystal is destroyed when the polarons start to overlap).
In this range, although the polarons do not survive  beyond the
transition, the resulting electron liquid should present both strong e-ph
and e-e interactions (region III). A second possibility
is that not all of the polarons are destroyed at the transition, because
this would cost too much potential energy, especially in the strong
coupling limit. This can be understood as follows: when the density of
\textit{localized} polarons is increased,
there is a competition between the increase of
$\langle \delta r^{2}\rangle ^{1/2}/R_s$ due to the zero-point
fluctuations (which destabilizes the PWC), and the tendency of the system to
preserve the potential energy stored in
the polaron bound states (which in turn should prevent dissociation).
A compromise could be achieved in a
hypothetical \textit{mixed phase}, where
some itinerant carriers could coexist with the localized polarons.
In fact, some experiments \cite{muller} suggest that all the carriers
introduced beyond the critical density are itinerant (their number being
proportional to
$n-n_c^{(ii)}$), the density of localized polarons being kept constant and
equal to $n_c^{(ii)}$. The properties of such a mixed phase are currently under
studies. It is clear, however, that this could be stable only close to
$n_c^{(ii)}$, while for $n\gg n_c^{(ii)}$, all the interactions would be
screened, and the system would tend to a normal Fermi liquid.

%

\subsection{Transition at $T\neq 0$}

If we  go to finite temperatures, the melting of the polaron crystal
can be studied in terms of the same Lindemann criteria $(i)$ and $(ii)$
as in the zero temperature case,
provided that we include the effect of thermal fluctuations in the
definition of $\langle \delta R^2 \rangle$ and $\langle \delta r^2
\rangle$. The correct result is given in the appendix (see eq.
(\ref{ratio1}) and (\ref{ratio2})), where
 the variational parameters $\alpha_1$ and $\alpha_2$ are  now
 obtained by minimizing the
free energy $F$ given by eq. (\ref{Var-F-bis}), together with (\ref{F_0}),
(\ref{Bfinal}) and (\ref{Afinal}). It is known from the single
polaron case, however, that the upper bound (\ref{Var-F}) that one
estimates starting from
the two-body model (\ref{L_0}) only gives correct results for $T<\omega_{LO}$
\cite{castrignano,FratPhD}. At higher temperatures, such quantities
as the polaron mass $M_P$ and internal frequency $\alpha_2$ are
 ill-defined, and the treatment is no longer applicable in this
simple form.
Nevertheless, since the characteristic energy scale for
the thermal fluctuations of the polaron center-of-mass is $\alpha_1
<\omega_{LO}$, we reasonably expect the thermal melting of the crystal to
occur at
temperatures well beyond the limits of validity of the treatment, as
 can be verified \textit{a posteriori}.

In order to understand the basic physics of the problem, we first restrict
ourselves to
the low-density regime as was done in section \ref{sec-lowdens}. This
allows us to express $\langle \delta R^2\rangle$ and $\langle \delta
r^2\rangle$ by the
approximate formulas
\begin{eqnarray}
        \langle \delta R^2\rangle & = & (2M_P \alpha_1)^{-1} \coth
\frac{\beta\alpha_1}{2}
         \label{rsimp1}\\
         \langle \delta r^2\rangle & = & (2\mu\alpha_2)^{-1} \coth
\frac{\beta\alpha_2}{2}
          \label{rsimp2}.
\end{eqnarray}
Polaron theory
predicts that at temperatures much lower than $\omega_{LO}$, the properties
of each polaron
are almost unaffected by thermal fluctuations
\cite{castrignano,schultz}. In that case, we can
consider to a first approximation $\alpha_2(T)\approx \alpha_2$,
$M_P(T)\approx M_P$, and consequently $\alpha_1(T)\approx \alpha_1$
(from eq. (\ref{relation(alpha)})). Therefore, in this simple limit, the
only temperature
dependence of the radii (\ref{rsimp1}) and (\ref{rsimp2}) comes from the
explicit $\beta$ factor.

In the intermediate and strong coupling regimes (i.e. where the mechanism
of polaron dissociation can become relevant), since $\alpha_2 \gg
\omega_{LO}$, $\coth (\beta\alpha_2/2)\approx 1$,
and we immediately see that the dissociation density does not
depend on the temperature:
\begin{equation}
        r_c^{(ii)}(T)\approx r_c^{(ii)}{(T=0)}
\end{equation}
On the other hand, criterion $(i)$ can be solved to give
the critical temperature for crystal melting:
\begin{equation}
        {T_c}=\omega_{ext}/\ln \left\lbrack
        \left({\sqrt{\displaystyle\frac{r_s}{r_c^{(i)}{\scriptstyle (T=0)}}}+1}\right)/
        \left({\sqrt{\displaystyle\frac{r_s}{r_c^{(i)}{\scriptstyle (T=0)}}}-1}\right)
        \right\rbrack
        \label{Tfusion}
\end{equation}
At low temperatures, this equation has two different solutions which
correspond respectively to the quantum melting analysed in section
\ref{sec:LindT0}, at $r_s\approx r_c^{(i)}{(T=0)}$, and to the classical
melting due to thermal fluctuations, at $r_s \rightarrow \infty$.
Let us first expand eq. (\ref{Tfusion}) in the classical
case:
\begin{equation}
        T_c \simeq \frac{2Ry}{\varepsilon_s r_s}\delta^2 \propto n^{1/3}
        \label{Tlow}
\end{equation}
The melting temperature in that case is proportional to the only relevant
energy scale in the problem, i.e. the strength of the Coulomb
interactions \cite{remark-MC}. In the quantum limit, i.e.
for $r_s$
close to the zero temperature value $r_c^{(i)}{ (T=0)}$,
 $T_c$ can be expressed as:
\begin{equation}
        T_c \simeq -{\omega_{ext}}/{\ln \left\lbrack
        \displaystyle
        \frac{1}{4}\frac{r_s-r_c^{(i)}{\scriptstyle (T=0)}}{r_c^{(i)}{\scriptstyle
(T=0)}}\right\rbrack}
        \label{Thigh}
\end{equation}
which states that the critical temperature tends to zero at $r_c^{(i)}{ (T=0)}$
with a  slope that diverges logarithmically.

\begin{figure}[ht]
        \centerline{\resizebox{8cm}{!}{\includegraphics{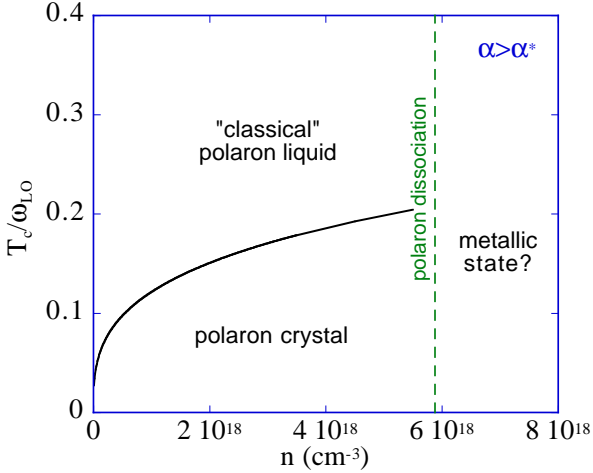}}}
        \caption{The $T$ vs. $n$ phase diagram for $\alpha=10>\alpha^*$, as obtained
from criteria $(i)$ and $(ii)$.} \label{diag-nT-sc}
\end{figure}

\begin{figure}[ht]
        \centerline{\resizebox{8cm}{!}{\includegraphics{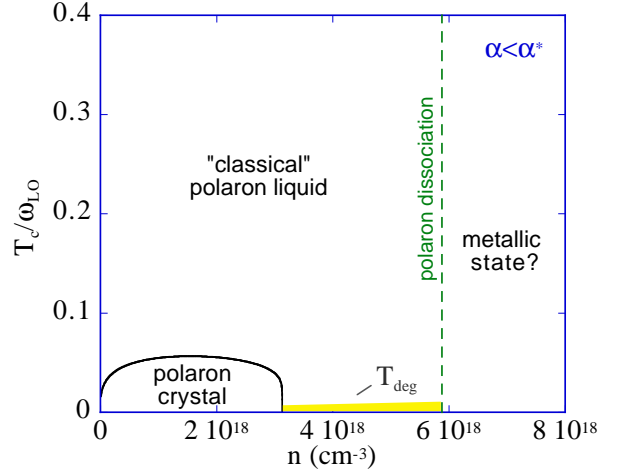}}}
        \caption{The $T$ vs. $n$ phase diagram for $\alpha=7.2 <\alpha^*$, as obtained
from criteria $(i)$ and $(ii)$. The shaded
area corresponds to temperatures $T<T_{deg}$, where the polaron liquid
could become degenerate.} \label{diag-nT-wc}
\end{figure}

In the general case (i.e. at any density and any e-ph coupling), the
Lindemann criteria $(i)$ and $(ii)$ can be accurately evaluated as was
indicated at the beginning of
this section. In figures \ref{diag-nT-sc} and \ref{diag-nT-wc}, we show
two characteristic phase diagrams in the
$(T,n)$ plane, which correspond respectively to e-ph coupling constants
above and below $\alpha^*$.
In the strong coupling regime, where $n_c^{(i)}>n_c^{(ii)}$
(Fig.\ref{diag-nT-sc}, $\alpha=10$), while
quantum fluctuations lead to
polaron dissociation, thermal fluctuations lead to the
melting through criterion $(i)$, so that a polaron liquid state can be achieved
by increasing the temperature above $T_c$.
However,
a simple argument shows that it is very unlikely that such
polaron liquid is degenerate.
In fact, the particle statistics become relevant when
the de Broglie thermal length  is comparable to their average distance, i.e.:
\begin{equation}
        \left(\frac{2\hbar^2}{M_Pk_BT}\right)^{1/2}\approx 2R_s
\end{equation}
Therefore, the polaron liquid is expected to behave classically down to
temperatures of the order
\begin{equation}
        T_{deg}\approx \frac{m_e}{M_P}\frac{2 \cdot 10^5 K}{r_s^2}
        \label{}
\end{equation}
If we consider a polaron mass $M_P\approx 10m_e$ and a density $r_s=100$, the
resulting $T_{deg}$ is of a few Kelvin, while $T_c$ can be as high as
some fraction of $\omega_{LO}$ (typically $\sim 100 K$).
In conclusion,
the phase diagram at strong coupling separates into
three different regions:  polaron crystal (at low density, low
temperature),  "classical" polaron liquid (at low density, high
temperature), and some liquid state at high density where the polarons (or
at least some of them) are ionized.

Let us now analyse the situation  for $\alpha$ lower than $\alpha^*$,
where $n_c^{(i)}<n_c^{(ii)}$ (Fig.\ref{diag-nT-wc}, $\alpha=7.2$). In
that case, by increasing either the density or the temperature, the crystal
melting
is always driven by
criterion $(i)$. Again, the phase diagram is divided in three regions,
but now the polaron liquid can  become degenerate at sufficiently
low temperature (i.e. for $T<T_{deg}$), in the
intermediate region between $n_c^{(i)}$ and $n_c^{(ii)}$ (shaded
area in Fig.\ref{diag-nT-wc}).

\subsection{Extension to the anisotropic case}
\label{sec:anisotropic}

Both the static ($\varepsilon _{s}$) and high frequency ($\varepsilon
_{\infty}$) dielectric constants have been measured in some undopped
cuprates. For the parent compound La$_2$CuO$_4$, it was found \cite{chen}
$\varepsilon _{s}\approx 30$ and $\varepsilon _{\infty} \approx 5$.
Surprisingly, the static dielectric constant remains almost {\it isotropic}
in the three directions of space (c-axis, and ab-plane of cuprates), which
means
that the phonons seem to act in the same manner in these three directions.
However, it is known that the transport properties are incoherent in the
c-direction of CuO$_2$ plane. A possible way to simulate such
a specific situation, is to consider a system where the electron bare band mass
$m^*$ is highly anisotropic, while the  Coulomb interactions (and the
electron-phonon interaction) remain isotropic.
This problem can  be studied by considering a
crystallized state of Fr\"{o}hlich polarons, where the localizing potential
in the Wigner approximation is
spherically symmetric, but the electron effective band masses
satisfy $b=m_{xy}/m_{z}\ll 1$.

If we assume $b\rightarrow 0$, which is equivalent to
confining the carriers in 2D
layers, it is known that the polaron bound states become
anisotropic (in this limit, they are usually called ''surface''
polarons \cite{sak}). Correspondingly, the expression (\ref{E_var})
for the variational energy at zero temperature must be replaced by:
\begin{eqnarray}
E &=&C_0+ \left( \alpha _{1}+\alpha _{2}-w\right) -
\frac{ \left( w^{2}-\alpha
_{1}^{2}\right) \left(\alpha _{2}^{2}-w^2 \right)  }{2\left(
\alpha _{1}+\alpha _{2}\right) w^{2}}  \nonumber \\
&-&\frac{\alpha }{\sqrt{\pi }} \frac{\pi}{2}\int_0^\infty
\frac{e^{-t}\,dt}{\sqrt{\frac{A_{1}}{%
\alpha _{1}}\left( 1-e^{-\alpha _{1}t}\right) +\frac{A_{2}}{\alpha _{2}}%
\left( 1-e^{-\alpha _{2}t}\right) }}   \label{energy2D}
\end{eqnarray}
If we compare this expression with (\ref{E_var}), we see that apart from
a numerical coefficient, the function to be minimized in 2D is equivalent
to the one in the 3D case, provided that we multiply $\alpha$ by a factor
$3\pi/4$
(such a scaling relation was first derived by  Peeters et al.
\cite{peeters} in the case of an isolated polaron). In our case, it is
easy to see that, at a given density, the variational parameters in 2D are
given by the
simple relations
\begin{eqnarray}
        \alpha_2^{2D}(\alpha)=\alpha_2^{3D}(3\pi\alpha/4) \nonumber\\
        \alpha_1^{2D}(\alpha)=\alpha_1^{3D}(3\pi\alpha/4) \nonumber
\end{eqnarray}
so that the phase diagram in the anisotropic case can be easily deduced
from the
3D results (see Fig.\ref{fig-phasediag}, grey lines \cite{remark-2D3D}).
The main conclusions are
summarized in Table 1, for two different values of the
electron band mass.
\begin{table}[ht]
    \begin{tabular}{l|ccc}
\hline \hline
{} & $\alpha^*$ & $n_c ^{(ii)}  (m^*=m_e)$ & $n_c^{(ii)} (m^*=2m_e)$ \\
\hline
isotropic (3D)   & $7.5$ & $6 \cdot 10^{18}$ & $ 5 \cdot 10^{19}$\\
anisotropic (2D) & $3.2$ & $8 \cdot 10^{19}$ & $ 6 \cdot 10^{20}$\\
\hline \hline
\end{tabular}
        \caption{Critical density $n_c^{(ii)} (cm^{-3})$
        and coupling $\alpha^*$ (see text) \cite{remark-anisotropy}.}
\end{table}
We first observe that the critical coupling above which
polaron dissociation becomes effective is
strongly reduced from $\alpha ^{*}\left( 3D\right) \approx 7.5$
(which corresponds to $\omega_{LO}\approx 6 meV$ if we take $m*=m_e$), to $
\alpha ^{*}\left( 2D\right) \approx 3.2$ ($\omega_{LO}\approx 40 meV$).
This is reminiscent of the fact that the e-ph coupling has a stronger
effect in lower
dimensions. Moreover, we see that the critical density $n_c^{(ii)}$ is
increased:
with the parameters of La$_2$CuO$_4$ ($m^*=2m_e$,
$\varepsilon_s=30$, $\varepsilon_\infty=5$ \cite{chen,LSCO}), one
obtains $n_{c}^{(ii)}\approx 6 \cdot 10^{20}cm^{-3}$, in qualitative
agreement with experiments. These results suggest that the mechanism of polaron
dissociation could be relevant in the description of the IMT
in the high-$T_{c}$ cuprates. Note however that it does not mean that the
electron motion would become coherent in the c-direction above the
dissociation. The resulting state above the dissociation is out of the
goal of the present paper.
%

\section{Conclusion}

In this work, we have attacked the problem of the insulator-to-metal
transition in doped polar semiconductors. Neglecting disorder, and
replacing the doping ions by a rigid compensating jellium, the two main
ingredients in the problem are: (i) the electron-phonon interaction, which
leads to polaron formation and (ii) the Coulomb repulsion between the
carriers, which causes their crystallization at low density. We have
studied the properties of the insulating crystallized state in the framework
of Wigner's mean-field approximation. The results show that, upon
increasing the Fr\"ohlich e-ph coupling $\alpha$, the Wigner
crystal of electrons evolves continuously towards a polaron crystal.
On the other hand, a definite distinction
between the weak and strong e-ph coupling limits comes from the melting mechanism
occurring for increasing density. By applying the Lindemann criterion both to
the polaron center-of-mass $\mathbf{R}$ and to the internal coordinate
$\mathbf{r}$,
we have shown
that, at $T=0$, the crystal melting due to the fluctuations of the localized
particles is only possible for $\alpha<\alpha^*$. In the opposite
situation $\alpha>\alpha^*$,
the IMT is driven by polaron dissociation:
the particles become more and more localized due to the increase in their
effective mass, and the only way to obtain a metallic state is to ionize
the polaron bound states.
We have also analysed the melting at finite temperature, and we have
proposed an approximate $T$ vs. $n$ phase diagram for the many-polaron
problem. Finally, we have shown that if the carriers are confined in 2D
layers, the effect of the e-ph interactions is strongly enhanced, so that
the mechanism of polaron dissociation can be relevant at rather moderate
couplings ($\alpha \approx 3$).

The overall picture which comes out from this work suggests that large
polarons should play an important role in the superconducting cuprates.
Moreover, it will be shown in a forthcoming paper \cite{frat2} by
fully taking into account the dipole-dipole interactions between polarons that,
due to the peculiar
dielectric properties of the PWC, if some itinerant carriers
could coexist with localized polarons, the long-range part of their
mutual interactions would be overscreened. This would lead to a
superconducting instability rather than to an insulator-to-metal transition, as
was already pointed out in reference \cite{frat1}.

It is also interesting to compare our results on the melting of a
PWC to the studies carried out by De Filippis et al.  for the same
model, but in the metallic phase \cite{iadonisi}.  These authors found that at
intermediate e-ph coupling ($\alpha \approx 6$), the metallic phase can
become unstable with respect to the formation of a Charge Density Wave
(CDW), as the density is decreased down to $n \approx 10^{18-20} cm^{-3}$.  For
the same value of $\alpha$, we find that in the same range of
densities, the  PWC becomes unstable
with respect to the formation of a polaron liquid.
In this sense, both studies on the same problem, either starting
from the low density limit or from the high density metallic phase,
seem to be quite consistent.
Nevertheless, our analysis suggests that the melting of the
 PWC at zero temperature is first order, as is the case for the Mott
transiton (the free carrier
density is a discontinuous quantity at the transition).  On the contrary,
the occurence of a CDW instability should rather be second order
 (the amplitude of the component $\rho_q$ of the electronic
density with CDW-wavevector $q$ being the order parameter).
We believe that at stronger coupling, where the polaron dissociation holds,
there should exist a \textit{metastability} of the metallic phase with
respect to the  PWC. This could be related to the fact that
the long-range Coulomb interactions (i.e. both
the electron-electron \textit{and}  the electron-phonon
interactions)  are completely unscreened in the crystallized state.

The behaviour of the long-range Coulomb
forces being an essential ingredient to the
understanding of this "intermediate density" physics is also suggested
by a recent work by Leggett \cite{leggett}. Arguing, on the basis of
experimental data on the cuprates, that the
Coulomb energy at the metal-superconducting transition
is saved in the mid infra-red range at small $q$ vectors,
Leggett has pointed out that the basic
pairing mechanism in high-Tc superconductors should also involve small
in-plane $q$ vectors, and thus could be due to long-range interactions.


Finally, we would like to mention that the present theoretical work can
also find applications in different classes of physical systems. In
particular, it was
shown by
Jackson and Platzman \cite{jackson} that the dynamics of ripplons coupled to
a 2D electron set on a film of liquid helium can be viewed as a
2D Fr\"ohlich polaron problem.

\smallskip

This work received financial support from the European Commission (contract
no. ERBFMBICT 961230).

\appendix
\section{Details of the Feynman treatment}

Here we evaluate the upper bound to the free energy $F$ according to eq. (\ref
{Var-F}), that we rewrite here as
\begin{equation}
        F=F_0+A+B
        \label{Var-F-bis}
\end{equation}

\subsection{Evaluation of the variational free energy $F$}

After the diagonalization has been carried out, the model (\ref{L_0})
represents
two independent harmonic oscillators in three dimensions. The
corresponding partition function is then
\begin{equation}
        Z_0=(2 \sinh \beta\alpha_1/2)^{-3} (2 \sinh \beta\alpha_2/2)^{-3}
        \label{partfunc}
\end{equation}
This result must be equivalent to the path integral of the electronic
action (\ref{qaction}), including the part corresponding to the motion of
the auxiliary
particle $\mathbf{X}$. In other words,
\begin{equation}
        Z_0=(2 \sinh \beta w/2)^{-3}\int \mathcal{D}(path) e^{S_0[\mathbf{u}]}
        \label{partbis}
\end{equation}
If we equate these two results, and make use of the definition
(\ref{path_energy}), we find:
\begin{equation}
        F_0=\frac{3}{\beta} \log \left\lbrack\frac{\sinh \beta w /2}
        {2\sinh \beta \alpha_1 /2 \sinh \beta \alpha_2/2} \right\rbrack
        \label{F_0}
\end{equation}

\bigskip

Let us now calculate the term $B=\left\langle S_0\right\rangle/\beta$,
which is defined as
\begin{equation}  \label{Bav}
B=-\frac{C}{2\beta} \int_0^\beta \int_0^\beta dt \, ds \, G_w(\beta,t-s)
\langle | \mathbf{ u}(t) - \mathbf{ u}(s) |^2 \rangle
\end{equation}
where $G_{w }(\beta ,t -s)$ is the phonon propagator defined in section
\ref{sec:phint}, and $C=Kw/4$.
The integrand in $B$ can be obtained by expanding up to 2nd order in $k_x$
the expression
\begin{equation}  \label{Ik1}
I(k_x,t,s)= \left\langle e^{ik_x \left[ x(t) - x(s) \right]} \right\rangle
\end{equation}
($x$ is one of the three equivalent cartesian coordinates:
$\mathbf{u}=(x,y,z)$).
If we write explicitly the path integral in the last expression and put
$f(\tau)=ik_x\delta(\tau-t) -ik_x\delta(\tau-s)$, eq. (\ref{Ik1}) takes the
form
\begin{eqnarray}
\lefteqn{I(k_x,t,s) \propto \int \mathcal{ D}(path) \, \exp \left \lbrack
-\frac{1}{2}
\int \dot{x}^2 d\tau -\frac{\omega_W^2}{2\varepsilon_s} \int x^2 d\tau
\right. } \nonumber \\
&-&   \frac{C}{2} \int \int |x(\tau) -
x(\sigma)|^2 G_w(\beta,\tau-\sigma) d\tau \, d\sigma  \nonumber \\
&+&\left.  \int f(\tau) \cdot x(\tau) d\tau \right\rbrack  \label{Gauss}
\end{eqnarray}
The calculation of such a gaussian integral is standard: first evaluate the
function $\bar{x}(\tau)$ for which the exponent is maximum, subject to the
boundary condition $\bar{x}(0)=\bar{x}(\beta)=0$, then (within an
unimportant constant) the integral (\ref{Gauss}) reduces to
\begin{eqnarray}
\lefteqn{I(k_x,t,s) = \exp \left \lbrack -\frac{1}{2} \int \dot{\bar{x}}^2
d\tau -%
\frac{\omega_W^2}{2\varepsilon_s} \int \bar{x}^2 d\tau \right. } \nonumber \\
&-& \frac{C}{2} \int\int |\bar{x}(\tau) - \bar{x}(\sigma)|^2
G_w(\beta,\tau-\sigma) d\tau \, d\sigma  \nonumber \\
&+& \left.  \int f(\tau) \, \bar{x}(\tau) d\tau \right \rbrack
\label{Gauss_class}
\end{eqnarray}
which can be simplified to
\[
I(k_x,t,s)= \exp \left\lbrace \frac{1}{2} \int f(\tau) \, \bar{x}(\tau)
d\tau \right\rbrace =e^{  \frac{ik_x}{2} \lbrack \bar{x}(t) -\bar{
x}(s) \rbrack }
\]
The function $\bar{x}(\tau)$ is the solution of the integral equation
\begin{equation}
\ddot{x}(\tau)=2C\int\lbrack x(\tau)-x(\sigma)\rbrack
e^{-w|\tau-\sigma|}d\sigma + \frac{\omega_{W}^{2}}{\varepsilon_s}
x(\tau)-f(\tau)  \label{diffeq}
\end{equation}
which can be solved by defining the auxiliary function
\begin{equation}
X(\tau)=\frac{w}{2}\int G_w(\beta,\tau-\sigma) x(\sigma) d\sigma  \label{auxil}
\end{equation}
Now the differential system reads
\begin{eqnarray}
\ddot{x}(t) & = & \frac{4C}{w}\lbrack x(t) - X(t) \rbrack
+\frac{\omega_W^2}{\varepsilon_s} x(t) -f(t) \\
\ddot{X}(t) & = & w^2 \lbrack X(t) - x(t) \rbrack  \nonumber
\end{eqnarray}
In fact, the problem
is equivalent to solving the equations of motion of the
two-body model (\ref{L_0})
under the external driving force $f(\tau)$. After some lenghty algebra, we find
\begin{equation}  \label{Ik2}
I(k_x,t,s)= \exp \left \lbrace - \frac{k_x^2}{2} g(t-s )\right\rbrace
\end{equation}
with
\begin{equation}
        g(t-s)=\frac{A_1}{\alpha_1}g_1(t-s)+\frac{A_2}{\alpha_2}g_2(t-s)
\end{equation}
and
\begin{equation}
        g_i(t-s)=\frac{\cosh \beta \alpha_i/2-\cosh \alpha_i
        (t-s-\beta/2)}{\sinh \beta\alpha_i} \; \; ; \; \; i=1,2
\end{equation}
In the isotropic case, by expanding to second order in $k$ both equations (%
\ref{Ik1}) and (\ref{Ik2}), with $k^2=k_x^2+k_y^2+k_z^2$ (the three
directions are equivalent) and performing the time integrations in eq. (\ref
{Bav}) we obtain
\begin{equation}
        B=-\frac{3C}{w(\alpha_2^2-\alpha_1^2)} \left\lbrack
        \alpha_2 \coth \frac{\beta\alpha_2}{2} -\alpha_1 \coth \frac{\beta \alpha_1}{2}
        \right\rbrack
        \label{Bfinal}
\end{equation}

\bigskip

The term $A=-\left\langle S \right\rangle/\beta$ is defined as
\begin{equation}
A=\frac{\alpha}{\sqrt{8}\beta} \int_0^\beta \int_0^\beta \left\langle |
{\mathbf{ u}(t) - \mathbf{ u}(s)
|^{-1} }\right\rangle G_{\omega_{LO}}(\beta,t-s) \, dt \, ds
\end{equation}
It can be calculated by introducing  the Fourier transform
\begin{eqnarray}
\langle | {\mathbf{ u}(t) - \mathbf{ u}(s) |^{-1} \rangle= \int
\frac{d^3k}{2\pi^2
k^2} \langle e^{i \mathbf{k} \cdot \left[ \mathbf{u}(t) - \mathbf{u}(s)
\right]} \rangle} \\
=\int \frac{d^3k}{2\pi^2 k^2} I(k_x,t-s) I(k_y,t-s) I(k_z,t-s)
\end{eqnarray}
The integration over $k$ gives the result
\begin{equation}
A=-\frac{\alpha}{\sqrt{\pi}} \frac{1}{1-e^{-\beta}}\int_0^\beta  \frac{dt
\, e^{-t}}{\sqrt{%
g(t)}} \label{Afinal}
\end{equation}

%
%

%
%
If we take the limit $\beta \rightarrow \infty$, and include the constant
part $C_0=-(9\tilde{\varepsilon}^2\alpha^2/5\varepsilon_sr_s)(m_e/m^*)$, the
upper bound to the
ground-state  energy takes the form
\begin{eqnarray}
E &=&C_0 +%
\frac{3}{2}(\alpha _{1}+\alpha _{2}-w)
-\frac{3(w^{2}-\alpha
_{1}^{2})(\alpha _{2}^{2}-w^{2})}{4(\alpha _{1}+\alpha _{2})w^{2}}   \nonumber
\\ &-&\frac{\alpha }{\sqrt{\pi }}\int_0^\infty
\frac{e^{-t}\,dt}{\sqrt{\frac{A_{1}}{%
\alpha _{1}}\left( 1-e^{-\alpha _{1}t}\right) +\frac{A_{2}}{\alpha _{2}}%
\left( 1-e^{-\alpha _{2}t}\right) }}  \label{E_var_app}
\end{eqnarray}

\subsection{Evaluation of the internal and external fluctuations}

Here we calculate the average displacements of the center-of-mass and of
the relative coordinate, which enter in the Lindemann criteria
$(i)$ and $(ii)$. In each space direction, since the
eigenmodes are independent,  we have $\langle u_{1}u_{2}\rangle =\langle
u_{1}\rangle
\langle u_{2}\rangle $ so that making use of eq. (\ref
{coord}) and (\ref{limit(mode)}), we get the dimentionless quantities
\begin{eqnarray}
\frac{\langle \delta R^{2}\rangle }{(2m^{*}\omega _{LO})^{-1}}
&=& \left(\frac{m^*}{M_P}\right)^{2}
\left[ A_{1}^{2} \frac{\alpha _{2}^{4}}{w^{4}} \;
\langle \delta u_{1}^{2}\rangle +A_{2}^{2} \frac{\alpha _{1}^{4}}{%
w^{4}} \; \langle \delta u_{2}^{2}\rangle \right]  \nonumber
\\
\frac{\langle \delta r^{2}\rangle }{(2m^{*}\omega _{LO})^{-1}} &=&  \left(
\frac{\alpha _{1}^{2}}{\alpha _{2}^{2}-\alpha _{1}^{2}}\right) ^{2}\langle
\delta u_{1}^{2}\rangle +\left( \frac{\alpha _{2}^{2}}{\alpha
_{2}^{2}-\alpha _{1}^{2}}\right) ^{2}\langle \delta u_{2}^{2}\rangle \nonumber
\end{eqnarray}
where
\begin{eqnarray}
\langle \delta u_{1}^{2}\rangle  &=&(2A_{1}\alpha _{1})^{-1} \coth \frac{\beta
\alpha_1}{2}\\
\langle \delta u_{2}^{2}\rangle  &=&(2A_{2}\alpha _{2})^{-1}\coth \frac{\beta
\alpha_2}{2}
\end{eqnarray}
are the r.m.s. displacements of two harmonic oscillators at finite temperature.
Observing that the polaronic length unit is
$(2m^*\omega_{LO})^{-1/2}=(m/m^*)\tilde{\varepsilon}\alpha a_0$, we can write
\begin{eqnarray} \label{ratio1}
\frac{\sqrt{\langle \delta R^{2}\rangle}}{R_{s}} &=&\left( \frac{\omega
_{W}^{2}}{4\alpha }\right) ^{1/3}%
\frac{m^*}{M_P} \times  \\
&  \times & \sqrt{\frac{A_{1}}{\alpha _{1}}\left( \frac{\alpha _{2}}{w}%
\right) ^{4}\coth \frac{\beta
\alpha_1}{2}+\frac{A_{2}}{\alpha _{2}}\left( \frac{\alpha _{1}}{w}\right)
^{4}\coth \frac{\beta
\alpha_2}{2}}\nonumber \\ \label{ratio2}
\frac{\sqrt{\langle \delta r^{2}\rangle}}{R_{s}} &=&
\left( \frac{\omega _{W}^{2}}{4\alpha }\right) ^{1/3}
\frac{1}{\alpha _{2}^{2}-\alpha _{1}^{2}} \times \\
&  \times &\sqrt{\frac{\alpha _{1}^{3}}{A_{1}}
\coth \frac{\beta
\alpha_1}{2}+
\frac{\alpha _{2}^{2}}{A_{2}}\coth \frac{\beta
\alpha_2}{2}} \nonumber
\end{eqnarray}


\begin{thebibliography}{}
\bibitem{remark-external} We call "external" any doping charge added to the
host material.
\bibitem{remark-notations} In refs. \cite{quem1,quem2,frat1}, the
notation $\omega_{ext}, \omega_{int}$ was used instead of
$\alpha_1,\alpha_2$ for the eigenfrequencies of the two-body model.
\bibitem{remark-MC} The correct value of $\delta$ for classical melting is
$0.155$, as can be deduced from MC simulations \cite{MC3D}.
\bibitem{remark-2D3D} In the limit
of vanishing e-ph coupling,
the melting density appears to be independent on the dimensionality.
This is due to the fact that we have considered the same
value of $\delta=0.25$ both in the 3D and in the 2D case. As a
consequence, the critical density estimated in the anisotropic case
should be considered as a lower bound.
\bibitem{remark-anisotropy} In previous references,  the
value $\alpha^*=4.4$ that we referred to as "fully anisotropic",
was in fact obtained by taking a small but finite value $b\approx 0.01$.
\end{thebibliography}
\end{document}